\documentclass[
,reprint % preprint / reprint
]{revtex4-1}

\usepackage[utf8]{inputenc}     % Umlaute (Linux)
\usepackage[T1]{fontenc}        % Worttrennung von Worten mit Umlauten möglich

\usepackage{amsmath}

\usepackage{graphicx}

\usepackage[squaren]{SIunits}
\newcommand{\SI}[2]{\unit{#1}{#2}}
\newcommand{\percent}[0]{\%}

\usepackage{color}              % Support for colors :\textcolor{red}{Text}
\usepackage{xspace}             % \xspace can be used in newcommands without parameters to ensure whitespace after it

\usepackage{ifthen}

\usepackage[
pdftex % hypertex (default), pdftex
,unicode % Unicode encoded pdf strings
,final, % final, draft, debug -> more text in log file
,pdfstartview={Fit}
,pdfpagelayout={SinglePage} % SinglePage, OneColumn, TwoPageLeft, TwoPageRight
,bookmarks={true} % generate bookmarks
,bookmarksopen={true} % open bookmarks tree with depth=<bookmarksopenlevel>
,bookmarksopenlevel=2
,bookmarksnumbered={true}
]{hyperref}

\begin{document}

\newcommand{\X}[1]{%
#1%
}

\newcommand{\todocite}  [0] {%
\todof{Reference to be added}
}
\newcommand{\todo}[1]{%
\textcolor{red}{TODO: #1}\xspace%
}
\newcommand{\todof}[1]{%
\textcolor{red}{\footnote{\textcolor{red}{TODO: #1}}}%
}%
\newcommand{\wort}[1]{%
\textcolor{violet}{#1}%
}
\newcommand{\klau}[1]{\textcolor{blue}{#1}}
\newcommand{\klauPd}[1]{\textcolor{magenta}{#1}}

\newcommand{\todoplot}[2]{%
\begin{figure}%
\centering%
\includegraphics{fig/dummy.pdf}%
\caption{\todo{#2}}%
\label{fig:#1}%
\end{figure}%
}
\renewcommand{\todoplot}[2]{\todo{Plot #2}}

\newcommand{\was}[1]{%
\textbf{#1}\\%
}

\newcommand{\intro}[1]{%
\textit{#1}%
}

\newcommand{\entdecker}[4][\empty]%
{\footnote{%
after #3 scientist \name{#2}%
\ifthenelse{\equal{#1}{\empty}}%
{}%
{\cite{#1}}%
\,(#4)%(1900-1950)
}}

\newcommand{\name}[1] {#1}% \textsc

\let\exporg\exp % damit keine endlosschleife erzeugt wird
\renewcommand{\exp}[1]{\exporg{\left(#1\right)}}

\newcommand{\K}[1] {\SI{#1}{\kelvin}}
\newcommand{\grad}[1] {\SI{#1}{\celsius}}
\newcommand{\nm}  [1] {\SI{#1}{\nano\meter}}
\newcommand{\cmVs}  [1] {\SI{#1}{\centi\meter\squared\per\volt\second}}
\newcommand{\um}  [1] {\SI{#1}{\micro\meter}}
\newcommand{\mm}  [1] {\SI{#1}{\milli\meter}}
\newcommand{\meV} [1] {\SI{#1}{\milli\electronvolt}}
\newcommand{\eV}  [1] {\SI{#1}{\electronvolt}}
\newcommand{\mV}  [1] {\SI{#1}{\milli\volt}}
\newcommand{\pa}[1] {\SI{#1}{\pascal}} % E-7 mbar = E-5 Pa
\newcommand{\uVK} [1] {\SI{#1}{\micro\volt\per\kelvin}} % \micro -> Font eurm10 at 10pt not found
\newcommand{\Scm} [1] {\SI{#1}{\siemens\per\centi\meter}}
\newcommand{\mr}[1] {\SI{#1}{MR}}
\newcommand{\wt}[1] {\SI{#1}{wt\percent}}
\newcommand{\sek}[1] {\SI{#1}{\second}}
\newcommand{\gMole}[1] {\SI{#1}{\gram\per\mole}}

\let\uorg\u
\renewcommand{\u}[1] {\SI{#1}{\atomicmass}}

\newcommand{\T} [1][\empty]%
{%
\ifthenelse{\equal{#1}{\empty}}%
{\ensuremath{T}\xspace}%
{\mbox{\ensuremath{T=\grad{#1}}}}%
}

\newcommand{\Tm} [1][\empty]%
{%
\ifthenelse{\equal{#1}{\empty}}%
{\ensuremath{T_\text{m}}\xspace}%
{\mbox{\ensuremath{T_\text{m}=\grad{#1}}}}%
}

\newcommand{\Tg} [1][\empty]%
{%
\ifthenelse{\equal{#1}{\empty}}%
{\ensuremath{T_\text{g}}\xspace}%
{\mbox{\ensuremath{T_\text{g}=\grad{#1}}}}%
}
\newcommand{\Tdep} [1][\empty]%
{%
\ifthenelse{\equal{#1}{\empty}}%
{\ensuremath{T_\text{dep}}\xspace}%
{\mbox{\ensuremath{T_\text{dep}=\grad{#1}}}}%
}
\newcommand{\Tsub} [1][\empty]%
{%
\ifthenelse{\equal{#1}{\empty}}%
{\ensuremath{T_\text{sub}}\xspace}%
{\mbox{\ensuremath{T_\text{sub}=\grad{#1}}}}%
}
\newcommand{\V} [1][\empty]%
{%
\ifthenelse{\equal{#1}{\empty}}%
{\ensuremath{V}\xspace}%
{\mbox{\ensuremath{V=\SI{#1}{\volt}}}}%
}

\newcommand{\cLong} [0]{conductivity\xspace}
\newcommand{\cLongL} [0]{conductivity \ensuremath{\sigma}\xspace}
\newcommand{\cLongs} [0]{conductivities\xspace}
\newcommand{\cLongsL} [0]{conductivities \ensuremath{\sigma}\xspace}
\let\corg\c % Backup
\renewcommand{\c} [1][\empty]%
{%
\ifthenelse{\equal{#1}{\empty}}%
{\ensuremath{\sigma}\xspace}%
{\mbox{\ensuremath{\sigma=\Scm{#1}}}}%
}

\newcommand{\CLong} [0]{doping concentration\xspace}
\newcommand{\CLongL} [0]{doping concentration \ensuremath{C}\xspace}
\newcommand{\CLongs} [0]{doping concentrations\xspace}
\let\Corg\C % Backup
\renewcommand{\C} [1][\empty]%
{%
\ifthenelse{\equal{#1}{\empty}}%
{\ensuremath{C}\xspace}%
{\mbox{\ensuremath{C=\mr{#1}}}}%
}
\newcommand{\Ckl} [1]{\mbox{\ensuremath{C<\mr{#1}}}}
\newcommand{\Cgr} [1]{\mbox{\ensuremath{C>\mr{#1}}}}

\newcommand{\SC}  [0] {semiconductor\xspace}
\newcommand{\SCs} [0] {semiconductors\xspace}
\newcommand{\CSC} [0] {conventional semiconductor\xspace}
\newcommand{\CSCs}[0] {conventional semiconductors\xspace}
\newcommand{\OSC} [0] {organic semiconductor\xspace}
\newcommand{\OSCs}[0] {organic semiconductors\xspace}
\newcommand{\SMU} [0] {source measure unit\xspace}

\newcommand{\IE}  [0] {ionization energy\xspace}
\newcommand{\EA}  [0] {electron affinity\xspace}
\newcommand{\IEs}  [0] {ionization energies\xspace}
\newcommand{\EAs}  [0] {electron affinities\xspace}
\newcommand{\HOMO}[0] {highest occupied molecular orbital\xspace}
\newcommand{\LUMO}[0] {lowest unoccupied molecular orbital\xspace}

\newcommand{\insitu}[0] {in-situ\xspace}
\newcommand{\etal}[0] { et~al.\xspace}
\newcommand{\eg}[0]  {\mbox{e.g.}\xspace} % lat: exempli gratia: "zum Beispiel → z.B." (for example)
\newcommand{\ie}[0]  {\mbox{i.e.}\xspace} % lat: id est: "das heißt → d.h." (that is)
\newcommand{\etc}[0] {\mbox{etc.}\xspace} % lat: et cetera, "und so" (and so on)
\newcommand{\vs}[0] {vs.\ }

\newcommand{\mphOFET}[0]{\footnote{measured by Moritz Philipp Hein (IAPP) in an OFET geometry on SiO$_2$ substrate}\xspace}

\newcommand{\fFD}     [0] {\ensuremath{f_\text{FD}}\xspace}
\newcommand{\fFDLong} [0] {Fermi-Dirac distribution function\xspace}
\newcommand{\fB}      [0] {\ensuremath{f_\text{B}}\xspace}
\newcommand{\fBLong}  [0] {Boltzmann distribution function\xspace}
\newcommand{\dos}     [0] {\ensuremath{DOS}\xspace}
\newcommand{\dosLong} [0] {density of states\xspace}
\newcommand{\dosLongL} [0] {density of states \ensuremath{DOS}\xspace}
\newcommand{\dosLongs} [0] {densities of states\xspace}
\newcommand{\dosLongsL} [0] {densities of states \ensuremath{DOS}\xspace}

\newcommand{\Td}  [0] {\ensuremath{T_\text{d}}\xspace}
\newcommand{\n} [0] {\ensuremath{n}\xspace}
\newcommand{\nLong} [0] {density of free charge carriers\xspace}
\let\notequal\ne % \notequal -> != /usually \ne, but I need \ne for n_e
\renewcommand{\ne} [0] {\ensuremath{n_\text{e}}\xspace}
\newcommand{\neLL} [0] {\ensuremath{n_\text{e,LL}}\xspace}
\newcommand{\neLong} [0] {density of free electrons\xspace}
\newcommand{\neLongL} [0] {\neLong \ensuremath{n_\text{e}}\xspace}
\newcommand{\neLongs} [0] {densities of free electrons\xspace}
\newcommand{\neLongsL} [0] {\neLongs \ensuremath{n_\text{e}}\xspace}
\newcommand{\nh} [0] {\ensuremath{n_\text{h}}\xspace}
\newcommand{\nhLL} [0] {\ensuremath{n_\text{h,LL}}\xspace}
\newcommand{\nhLong} [0] {density of free holes\xspace}
\newcommand{\nhLongL} [0] {\nhLong \ensuremath{n_\text{h}}\xspace}
\newcommand{\nhLongs} [0] {densities of free holes\xspace}
\newcommand{\nhLongsL} [0] {\nhLongs \ensuremath{n_\text{h}}\xspace}
\newcommand{\neh} [0] {\ensuremath{n_\text{e/h}}\xspace}
\newcommand{\nehLL} [0] {\ensuremath{n_\text{e/h,LL}}\xspace}

\newcommand{\nM} [0] {\ensuremath{n_\text{Mol}}\xspace}
\newcommand{\nMLong} [0] {total density of molecules\xspace}
\newcommand{\nH} [0] {\ensuremath{n_\text{H}}\xspace}
\newcommand{\nHLong} [0] {density of host molecules\xspace}
\newcommand{\nD} [0] {\ensuremath{n_\text{D}}\xspace}
\newcommand{\nDLong} [0] {density of dopant molecules\xspace}
\newcommand{\nDLongL} [0] {density of dopant molecules \ensuremath{n_\text{D}}\xspace}

\newcommand{\DopEff} [1][\empty]{\ifthenelse{\equal{#1}{\empty}}%
 {\ensuremath{\eta_\text{dop}}\xspace}%
 {\mbox{\ensuremath{\eta_\text{dop}=\SI{#1}{\percent}}}}%
}
\newcommand{\DopEffLL} [1][\empty]{\ifthenelse{\equal{#1}{\empty}}%
 {\ensuremath{\eta_\text{dop,LL}}\xspace}%
 {\mbox{\ensuremath{\eta_\text{dop,LL}=\SI{#1}{\percent}}}}%
}
\newcommand{\DopEffLong} [0] {doping efficiency\xspace}
\newcommand{\DopEffLongs} [0] {doping efficiencies\xspace}
\newcommand{\DopEffLongL} [0] {\DopEffLong \ensuremath{\eta_\text{dop}}\xspace}

\let\weg\ni % was some math symbol
\renewcommand{\ni}[0]{\ensuremath{n_\text{i}}\xspace}

\newcommand{\EvLong} [0] {valence energy\xspace}
\newcommand{\Ev} [0] {\ensuremath{E_\text{V}}\xspace}
\newcommand{\EcLong} [0] {conduction energy\xspace}
\newcommand{\Ec} [0] {\ensuremath{E_\text{C}}\xspace}
\newcommand{\Egap} [1][\empty]{\ifthenelse{\equal{#1}{\empty}}%
 {\ensuremath{E_\text{gap}}\xspace}%
 {\mbox{\ensuremath{E_\text{gap}=\eV{#1}}}}% eV!!!
}
\newcommand{\Ed} [0] {\ensuremath{E_\text{D}}\xspace}
\newcommand{\Ea} [0] {\ensuremath{E_\text{A}}\xspace}
\newcommand{\EtLong} [0] {transport level\xspace}
\newcommand{\EtLongL} [0] {\EtLong \ensuremath{E_\text{Tr}}\xspace}
\newcommand{\Et} [1][\empty]{\ifthenelse{\equal{#1}{\empty}}%
 {\ensuremath{E_\text{Tr}}\xspace}%
 {\mbox{\ensuremath{E_\text{Tr}=\meV{#1}}}}%
}

\newcommand{\EfLong} [0] {Fermi level\xspace}
\newcommand{\EfLongL} [0] {\EfLong \ensuremath{E_\text{F}}\xspace}
\newcommand{\Ef} [0] {\ensuremath{E_\text{F}}\xspace}

\newcommand{\EsLong} [0] {Seebeck energy\xspace}
\newcommand{\EsLongL} [0] {\EsLong \ensuremath{E_\text{S}}\xspace}
\newcommand{\Es} [1][\empty]{\ifthenelse{\equal{#1}{\empty}}%
 {\ensuremath{E_\text{S}}\xspace}%
 {\mbox{\ensuremath{E_\text{S}=\meV{#1}}}}%
}

\newcommand{\EactLong} [0]{activation energy of the conductivity\xspace}
\newcommand{\EactLongL} [0]{\EactLong \ensuremath{E_{\text{act,}\c}}\xspace}
\newcommand{\EactLongs} [0]{activation energies of the conductivity\xspace}
\newcommand{\Eact} [1][\empty]{\ifthenelse{\equal{#1}{\empty}}%
 {\ensuremath{E_{\text{act,}\c}}\xspace}%
 {\mbox{\ensuremath{E_{\text{act,}\c}=\meV{#1}}}}%
}

\newcommand{\Nv} [0] {\ensuremath{N_\text{V}}\xspace}
\newcommand{\Nc} [0] {\ensuremath{N_\text{C}}\xspace}
\newcommand{\Nd} [0] {\ensuremath{N_\text{D}}\xspace}
\newcommand{\Ndi}[0] {\ensuremath{N_\text{D}^+}\xspace}
\newcommand{\Na} [0] {\ensuremath{N_\text{A}}\xspace}
\newcommand{\Nai}[0] {\ensuremath{N_\text{A}^-}\xspace}
\newcommand{\gd} [0] {\ensuremath{g_\text{D}}\xspace}
\newcommand{\ga} [0] {\ensuremath{g_\text{A}}\xspace}

\newcommand{\kB} [0] {\ensuremath{k_\text{B}}\xspace}
\newcommand{\kT} [0] {\ensuremath{k_\text{B}T}\xspace}

\newcommand{\EField} [0] {\ensuremath{\mathcal{E}}\xspace}

\newcommand{\dc} [0] {\ensuremath{d_\text{c}}\xspace} % contact distance
\newcommand{\lc} [0] {\ensuremath{l_\text{c}}\xspace} % contact length
\newcommand{\hl} [0] {\ensuremath{h_\text{l}}\xspace} % layer height / thickness

\newcommand{\mob}  [1][\empty]{\ifthenelse{\equal{#1}{\empty}}%
 {\ensuremath{\mu}\xspace}%
 {\mbox{\ensuremath{\mu=\cmVs{#1}}}}%
}
\newcommand{\mobLL}  [1][\empty]{\ifthenelse{\equal{#1}{\empty}}%
 {\ensuremath{\mu_\text{LL}}\xspace}%
 {\mbox{\ensuremath{\mu_\text{LL}=\cmVs{#1}}}}%
}
\newcommand{\mobUL}  [1][\empty]{\ifthenelse{\equal{#1}{\empty}}%
 {\ensuremath{\mu_\text{UL}}\xspace}%
 {\mbox{\ensuremath{\mu_\text{UL}=\cmVs{#1}}}}%
}
\newcommand{\mobe} [0] {\ensuremath{\mob_\text{e}}\xspace}
\newcommand{\mobh} [0] {\ensuremath{\mob_\text{h}}\xspace}
\newcommand{\mobeh} [0] {\ensuremath{\mob_\text{e/h}}\xspace}

\newcommand{\Vs} [0] {\ensuremath{V_\text{S}}\xspace} % Seebeck Voltage
\let\Sorg\S % \S was §
\newcommand{\SLong}[0]{Seebeck coefficient\xspace}
\newcommand{\SLongL}[0]{\SLong \ensuremath{S}\xspace}
\newcommand{\SLongs}[0]{Seebeck coefficients\xspace}
\renewcommand{\S} [1][\empty]{\ifthenelse{\equal{#1}{\empty}}%
 {\ensuremath{S}\xspace}%
 {\mbox{\ensuremath{S=\uVK{#1}}}}%
}
\newcommand{\MM}  [0] {\ensuremath{M}\xspace} % Molar Mass
\newcommand{\density}  [0] {\ensuremath{\rho}\xspace}

\newcommand{\gausswidth} [1][\empty]{\ifthenelse{\equal{#1}{\empty}}%
 {\ensuremath{\sigma_\text{G}}\xspace}%
 {\mbox{\ensuremath{\sigma_\text{G}=\meV{#1}}}}%
}

\newcommand{\gausscenter} [0] {\ensuremath{E_\text{G}}\xspace}
\newcommand{\peltier}     [0] {\ensuremath{\Pi}\xspace}
\newcommand{\avogadro}    [0] {\ensuremath{N_\text{Avo}}\xspace}
\newcommand{\rms}         [0] {\ensuremath{R_\text{rms}}\xspace}
\newcommand{\pKa}         [0] {pKa\xspace}
\newcommand{\druck}       [0] {\ensuremath{P}\xspace} % in Bildern verwendet, z.B. killing+reanimating-C60-2-evap

\newcommand{\meo} [0] {\mbox{MeO-TPD}\xspace}
\newcommand{\meoLong} [0] {N,N,N',N'-tetrakis 4-methoxyphenyl-benzidine\xspace}
\newcommand{\lili} [0] {\mbox{BF-DPB}\xspace}
\newcommand{\liliLong} [0] {N,N'-Bis(9,9-dimethyl-fluoren-2-yl)-N,N'-diphenyl-benzidine\xspace}
\newcommand{\pen} [0] {pentacene\xspace}
\newcommand{\bphenLong}[0]{4,7-diphenyl-1,10-phenanthroline\xspace}
\newcommand{\znpc} [0] {ZnPc\xspace}
\newcommand{\CSF} [0] {\texorpdfstring{C$_{60}$F$_{36}$}{C60F36}\xspace}
\newcommand{\FS} [0] {\texorpdfstring{\mbox{F$_{6}$-TCNNQ}}{F6-TCNNQ}\xspace}
\newcommand{\FSLong}[0] {1,3,4,5,7,8-hexafluorotetracyanonaphthoquinodimethane\xspace}
\newcommand{\FV} [0] {\texorpdfstring{\mbox{F$_{4}$-TCNQ}}{F4-TCNQ}\xspace}
\newcommand{\FVLong}[0] {tetrafluoro-tetracyanoquinodimethane\xspace}
\newcommand{\CS} [0] {\texorpdfstring{C$_{60}$}{C60}\xspace}
\newcommand{\CrPd}[0] {\texorpdfstring{Cr$_2$(hpp)$_4$}{Cr2(hpp)4)}\xspace}
\newcommand{\WPd}[0] {\texorpdfstring{W$_2$(hpp)$_4$}{W2(hpp)4)}\xspace}
\newcommand{\CrPdLong}[0] {tetrakis(1,3,4,6,7,8-hexahydro-2H-pyrimido[1,2-a]pyrimidinato)\-dichromium (II)\xspace}
\newcommand{\WPdLong}[0]  {tetrakis(1,3,4,6,7,8-hexahydro-2H-pyrimido[1,2-a]pyrimidinato)\-ditungsten (II)\xspace}
\newcommand{\aob} [0] {AOB\xspace}
\newcommand{\aobLong} [0] {3,6-bis(dimethylamino)acridine\xspace}
\newcommand{\dmbiPOH}[0] {\mbox{DMBI-POH}\xspace}
\newcommand{\dmbi}[0] {\dmbiPOH}
\newcommand{\dmbiPOHLong}[0] {2-(1,3-dimethyl-1\textsl{H}-benzoimidazol-3-ium-2-yl)phenolatehydrate\xspace}
\newcommand{\OHdmbi}[0] {\mbox{OH-DMBI}\xspace}
\newcommand{\OHdmbiLong}[0]  {2-(1,3-dimethyl-2,3-dihydro-1\textsl{H}-benzoimidazol-2-yl)phenol\xspace}
\newcommand{\meodmbiI}[0] {\mbox{\textsl{o}-MeO-DMBI-I}\xspace}
\newcommand{\meodmbiILong}[0] {2-(2-methoxyphenyl)-1,3-dimethyl-1\textsl{H}-benzoimidazol-3-ium iodide\xspace}
\newcommand{\meodmbi}[0] {\mbox{\textsl{o}-MeO-DMBI}\xspace}
\newcommand{\meodmbiLong}[0] {2-(2-methoxyphenyl)-1,3-dimethyl-2,3-dihydro-1\textsl{H}-benzoimidazol\xspace}
\newcommand{\Ndmbi}[0] {\mbox{N-DMBI}\xspace}
\newcommand{\NdmbiLong}[0]  {(4-(1,3-dimethyl-2,3-dihydro-1\textsl{H}-benzoimidazol-2-yl)phenyl)dimethylamine\xspace}
\newcommand{\PCBM}[0]  {PCBM\xspace}
\newcommand{\PCBMLong}[0]  {[6,6]-phenyl-C$_{61}$-butyric acid methyl ester\xspace}

\newcommand{\cBild}[3][tb]{
\begin{figure}[#1]% not h at first place!
\centering
\includegraphics[width=\columnwidth]{fig-#2.pdf}%  the '%' is important!!! , removed width [width=#3]
\caption{#3}% \caption[shortcaption]{caption}
\label{fig:#2}%
\end{figure}%
}

\newcommand{\PUNKT}[0]{\ensuremath{~.}}
\newcommand{\KOMMA}[0]{\ensuremath{~,}}

\newcommand{\sub}  [1] {\ensuremath{_\text{#1}}}

\newcommand{\Eqnref}[1] {Equation~(\ref{eq:#1})}
\newcommand{\eqnref}[1] {equation~(\ref{eq:#1})}
\newcommand{\eqnrefPage}[1] {\eqnref{#1} on page~\pageref{eq:#1}}
\renewcommand{\eqref}[1] {(\ref{eq:#1})}
\newcommand{\Figref}[1] {FIG.~\ref{fig:#1}}
\newcommand{\figref}[1] {FIG.~\ref{fig:#1}}
\newcommand{\secref}[1] {section~\ref{sec:#1}}
\newcommand{\Secref}[1] {Section~\ref{sec:#1}}
\newcommand{\tabref}[1] {table~\ref{tab:#1}}
\newcommand{\tabrefPage}[1] {table~\ref{tab:#1} on page~\pageref{tab:#1}}
\newcommand{\todosecref}[0] {section~\textcolor{red}X}

\newcommand{\lasteq}[1][0]{%
(%
\arabic{chapter}%
.%
\addtocounter{equation}{-#1}\arabic{equation}\addtocounter{equation}{#1}%
)\xspace%
}
\newcommand{\lasteqn}[1][0]{% same + word equation
equation~(%
\arabic{chapter}%
.%
\addtocounter{equation}{-#1}\arabic{equation}\addtocounter{equation}{#1}%
)\xspace%
}

\newcounter{tempCounter}
\let\theequationBackup\theequation

\newcommand*\cleartoleftpage{%
  \clearpage
  \ifodd\value{page}\hbox{}\thispagestyle{empty}\newpage\fi
}

\renewcommand{\todo}[1]{}

\newcommand{\affIapp}   [0] {Institut f\"ur Angewandte Photophysik, Technische Universit\"at Dresden, 01062 Dresden, Germany, http://www.iapp.de}
\newcommand{\affOxford} [0] {Physics Department, University of Oxford, Clarendon Laboratory, Parks Road, Oxford OX1 3PU, England, United Kingdom}
\newcommand{\affDebu} [0] {Department of Electrical Engineering, I.I.T. Madras, Chennai 600036, India}

\title{Determining doping efficiency and mobility from conductivity and Seebeck data of \texorpdfstring{\textit{n}-doped \boldmath\CS}{n-doped C60} layers}


\date{\today}
\begin{abstract}
\noindent
In this work, we introduce models for deriving lower limits for the key parameters doping efficiency, charge carrier concentration, and charge carrier mobility from conductivity data of doped organic semiconductors. The models are applied to data of thin layers of Fullerene \CS $n$-doped by four different $n$-dopants.
Combining these findings with thermoelectric Seebeck data, the energetic position of the transport level can be narrowed down and trends for the absolute values are derived.%
\end{abstract}

\author{Torben Menke}
\author{Debdutta Ray}
\thanks{Now at \affDebu}
\author{Hans Kleemann}
\author{Karl Leo}
\email{leo@iapp.de}
\author{Moritz Riede}
\thanks{Now at \affOxford}
\affiliation{\affIapp}

\maketitle

\section{Introduction}
The success of inorganic semiconductor devices relies to a large extend on the possibility to control the electrical conductivity and the position of the Fermi level within the semiconductor by electrochemical doping. 
Also in case of organic semiconductors, doping brings additional benefit to the device, making organic light-emitting diodes (OLEDs), organic photovolataic cells (OPV), and organic field-effect transistors (OFETs) more efficient and reliable.\cite{Reineke2009, Riede2011, Ante2011}
Similar to inorganic semiconductors, doping can be achieved for organic semiconductors by deploying either electron donating ($n$-type) or accepting ($p$-type) compounds (atoms or molecules).\cite{Walzer2007,LuessemRiedeLeo2013-PSS}
However, while doping in inorganic materials can be sufficiently described within the picture of one transport level and Fermi-Dirac statistics, the complexity of the description of doping in organic semiconductors is caused by their structural and energetical disorder.\cite{Gregg2004Review, Arkhipov2005, Mityashin2012a, Salzmann2012}
In particular, a self-consistent picture describing the complex interplay between the occupancy of the spatially and energetically distributed sites, the density dependence of charge carrier mobility and the number of ionized doping states has not been drawn. 

The aim of this work is to contribute to this understanding by presenting models for the estimation of key parameters of doped organic semiconductor layers: doping efficiency, free charge carrier concentration and mobility.
This work focuses on $n$-doping of small molecules, but the derived models \X{can be applied to} $p$-doping and polymers as well. 
First, lower limits for the electron mobility, the charge carrier density, as well as the doping efficiency are derived from conductivity data for $n$-doped \CS samples, comparing four different dopants. 
Second, combining these findings with thermoelectric Seebeck data, the energetic position of the transport level can be narrowed down. 
Finally, absolute values for the key parameters are derived, assuming a constant transport level.

\section{Materials and methods}
\begin{figure}%
\centering
\includegraphics{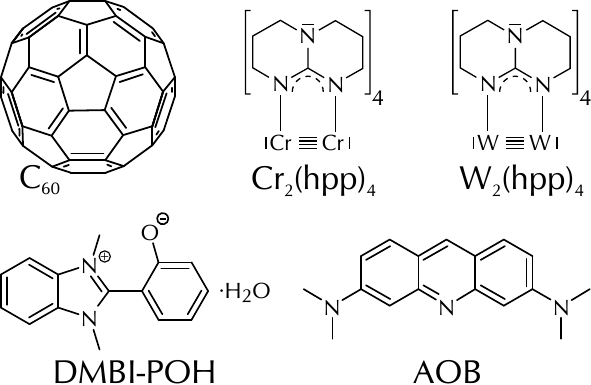}%
\caption{Chemical structures of the investigated materials.}
\label{fig:mat}%
\end{figure}
The prominent electron transporting material \CS is studied, comparing four $n$-dopants:
\CrPdLong and -ditungsten (II) (\CrPd\cite{Menke2012} and \WPd\cite{Menke2012}),
\aobLong (\aob\cite{Menke2012a}), and \dmbiPOHLong (\dmbiPOH\cite{Menke2012a}). The corresponding chemical structures are shown in \figref{mat}.
\CrPd and \WPd are rather heavy compounds with low \IE and hence reactive with air. \aob and \dmbiPOH are more light-weight, air-stable precursor compounds that form their active dopant compound during thermal co-deposition with \CS\cite{Menke2012a,Naab2013}.
The measurement data have been published earlier\cite{Menke2012,Menke2012a,TorbenMenkeDiss}, where details concerning sample fabrication and measurement techniques are given.
In this work, the data are re-evaluated to derive models and trends for the key parameters of doped layers, while comparing different dopants.

In the following, the \CLongL is expressed in terms of molar ratio MR being the ratio of the number densities $n$ of dopant (D) to host (H) molecules:
\begin{equation}
\C = \frac{\nD}{\nH}\label{eq:C}
\end{equation} 
The sum of host and dopant number densities gives the total number density of molecules \nM
\begin{equation}
\nM = \nH + \nD \label{eq:nMHD} 
\PUNKT
\end{equation}
From equations~\eqref{C} and \eqref{nMHD} follows
\begin{equation}
 \nD = \nM \cdot \frac{\C}{1+\C}  \label{eq:nD} 
\PUNKT
\end{equation} 

In an intrinsic layer, \nM can be derived from the material's mass density \density and molar mass \MM, together with Avogadro's constant \avogadro:
\begin{equation}
\nM = \frac{\avogadro \cdot \density}{\MM}
\label{eq:nM-from-MatPara}
\PUNKT
\end{equation} 
For \CS the values $\density=\SI{1.63}{\gram\per\centi\meter^{3}}$ and $\MM=\SI{720.6}{\gram\per\mole}$ yield a molecular density of \mbox{$\nM{}_{\text{,C}_{60}}=\SI{1.36\cdot10^{21}}{\centi\meter^{-3}}$}.
In the following, the common assumption is made that each dopant molecule substitutes one host molecule and hence \nM is unaltered upon doping. 

The electrical conductivity \c of an $n$-doped semiconductor can be expressed as the product of \neLongL and the electron mobility \mob
\begin{equation}\label{eq:conductivity}
\c = e \cdot \ne \cdot \mob
\KOMMA
\end{equation}
with the elementary charge $e$. 
In doped layers, \ne is proportional to the number \nDLongL and the doping efficiency \DopEff (neglecting the much lower intrinsic charge carrier and trap densities\cite{Pahner2013})
\begin{equation}
\label{eq:ne-from-DopEff-nD}
 \ne = \DopEff \cdot \nD
\PUNKT
\end{equation} 
\X{Inserting \eqnref{nD} into \eqref{ne-from-DopEff-nD}
and using \eqref{conductivity}, }
\ne can be correlated to \mob for a measured conductivity \c as
\begin{equation}
\mob = \frac{\c}{e \cdot \DopEff \cdot \nM} \cdot \frac{1+\C}{\C}
\label{eq:mob-von-DopEff+C}
\PUNKT
\end{equation} 
Both, \mob and \DopEff, are expected to vary with \CLong.

\section{Results}

\subsection{Measurement data}
\begin{figure}% not h at first place!
\centering
\includegraphics[width=\columnwidth]{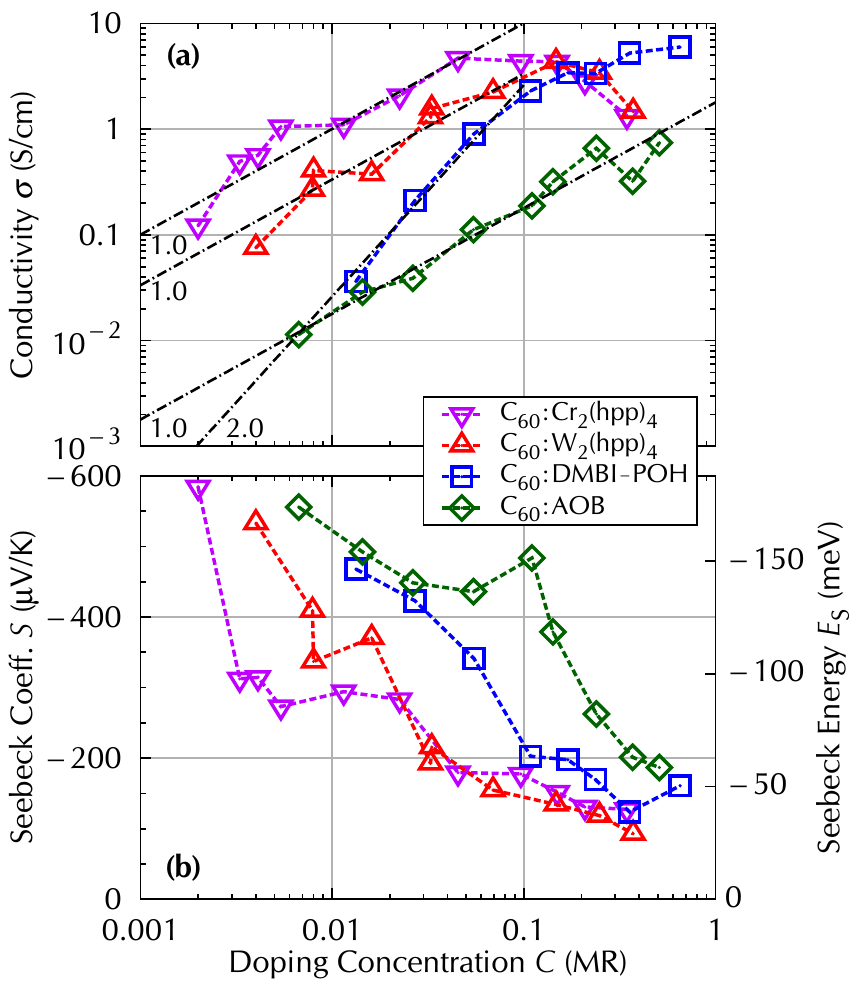}
\caption{%
Measurement data probed in vacuum at \T[40] (after thermal annealing) of 20 to \nm{30} thin \CS layers $n$-doped by air-stable (\aob and \dmbi) and air-sensitive (\CrPd and \WPd) dopants, combined from refs~\cite{Menke2012,Menke2012a,TorbenMenkeDiss}. 
(a) Conductivity \c and (b) Seebeck coefficient \S and Seebeck energy \Es \vs doping concentration \C. The chain-dotted lines of slopes 1.0 and 2.0 are guides to the eye.
}% \caption[shortcaption]{caption}
\label{fig:DataCond+See}%
\end{figure}%

Conductivity and thermoelectric Seebeck data, shown in \figref{DataCond+See},
allow for direct comparison of different $n$-doped \CS systems (20 to \nm{30} layer thickness).
Data are measured in vacuum at \T[40], after a thermal annealing step at \grad{100} that ensures reproducibility, as discussed in detail in the original publications.
Since all samples showed a linear and symmetric current--voltage dependency, contact resistances are neglected and ohmic injection is assumed.

\subsection{Lower limit of the mobility}\label{sec:rechMobLL}
\X{According to \eqnref{mob-von-DopEff+C}, \mob is inversely proportional to \DopEff for a given (measured) conductivity. Consequently, assuming a perfect doping efficiency of \DopEff[100] as upper limit (\ie each dopant molecule is ionized and provides one free charge carrier) a lower limit for the mobility~\mobLL can be derived from conductivity data at a given doping concentration:}
\begin{equation}
\mobLL = \frac{\c}{e \cdot 100\,\% \cdot \nM} \cdot \frac{1+\C}{\C}
\label{eq:mobLL}
\PUNKT
\end{equation}
\X{As in general the real \DopEff is below $100\,\percent$, the real mobility \mob must be higher than \mobLL to fulfill \eqnref{mob-von-DopEff+C}.}
Recently, it has been shown\cite{Olthof2012} that by using dimer molecules as dopants, two free electrons can be generated per dopant molecule. However, for the dopants used here such behavior is unlikely.

\cBild
{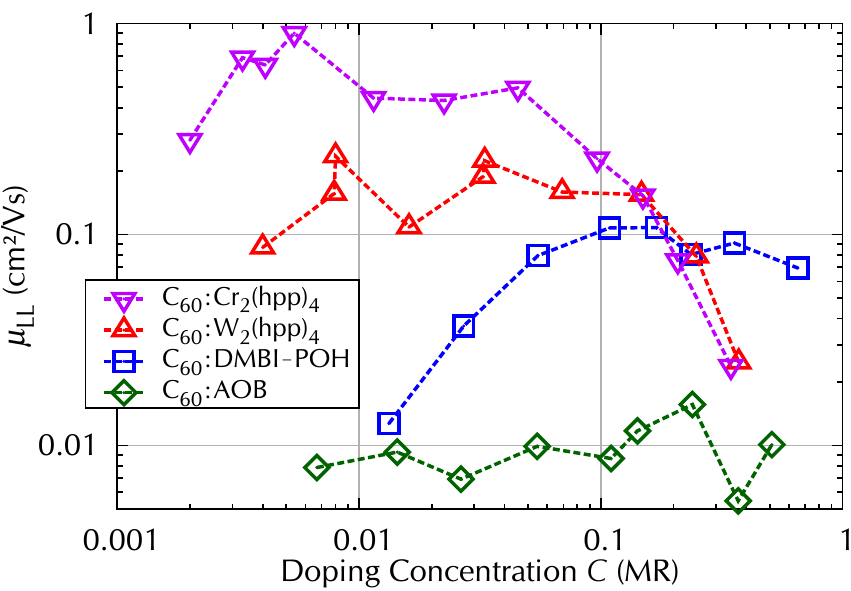}
{Lower limit of the electron mobility \mobLL for $n$-doped \CS, calculated by \eqnref{mobLL} using the conductivity data from \figref{DataCond+See}\,(a), probed at \T[40].}
In \figref{calc-MobLL-n}, this calculation is performed using the $n$-doped \CS conductivity data from \figref{DataCond+See}\,(a), probed at \T[40].
\mobLL is found to be highest for the dopant \CrPd with a maximum of \mobLL[0.9] at \C[0.005]. This value is close to the record mobility reported for undoped \CS layers of \mob[4.9]\cite{Itaka2006}, which is measured in an OFET geometry and thus at high \ne.
At \CLongs of \Ckl{0.045}, \mobLL is rather constant for \CrPd with values in the range of \mobLL[0.5], whereas at higher \C the \mobLL drops significantly.
A similar trend is found for the second air-sensitive dopant \WPd. At \Ckl{0.150}, almost constant values in the range of \mobLL[0.15] are derived, followed by a drop for higher \CLongs.
The reduction of the \mobLL suggest that the high density of the large and heavy dopants in the layer leads to a hindering of the transport and hence a reduction of the mobility, as discussed in ref.\cite{Menke2012} and supported by OFET studies on $n$-doped \CS layers\cite{Harada2007}.

A different relation is found for the more light-weight air-stable dopants \aob and \dmbi. The samples doped by \aob yield an almost constant value in the order of only \mobLL[9\cdot10^{-3}].
Samples doped by \dmbi start for low \C at a similar value to \aob, but show a strong increase with \C.
A saturation around \mobLL[0.1] is observed, being even higher than for the air-sensitive dopants at these \CLongs.

It is expected that for low \CLongs of each dopant the real values of the electron mobility are the same. Therefore, the different values of the \mobLL indicate different doping efficiencies of the dopants, which is addressed in the next section. 
From this model it cannot be distinguished whether the observed decrease of \mobLL at high \C for \CrPd and \WPd is correlated to trends of the real mobility \mob or a decreasing \DopEff.

\subsection{Lower limit of the charge carrier density and the doping efficiency}\label{sec:rechDopEffLL}
\X{Apart from estimations for the lower limit of the mobility by assuming \DopEff[100], the opposite approach can be performed by knowledge of an upper limit for the mobility \mobUL. 
Such an \mobUL allows for deriving a lower limit of the \neLong \neLL and doping efficiency \DopEffLL via equations~\eqref{conductivity} and \eqref{mob-von-DopEff+C}}
\begin{align}
\neLL &= \frac{\c}{e\cdot\mobUL} \label{eq:nehLL}\\
\DopEffLL &= \frac{\c}{e \cdot \mobUL \cdot \nM} \cdot \frac{1+\C}{\C} \label{eq:DopEffLL}
\PUNKT
\end{align}
As the real mobility in the used sample geometry is expected to be lower than this upper limit \mobUL and furthermore to be negatively affected by the introduction of dopant molecules hindering the transport, the real values of \ne and \DopEff must be larger than \neLL and \DopEffLL to fulfill \eqnref{mob-von-DopEff+C}.

The highest reported electron mobility in \CS is \mob[4.9]\cite{Itaka2006}, measured in an OFET geometry.
The free charge carriers in OFETs are generated by the electric field induced by the gate voltage and their number is typically much larger than values achieved by doping.\cite{Kaake2010} 
As the mobility of an \OSC usually increases with charge carrier density\cite{Pasveer2004}, OFET channel mobilities are generally larger than the mobilities in the bulk material and hence in the conductivity geometry. Therefore, this value can be interpreted as an upper limit \mobUL.

\cBild
{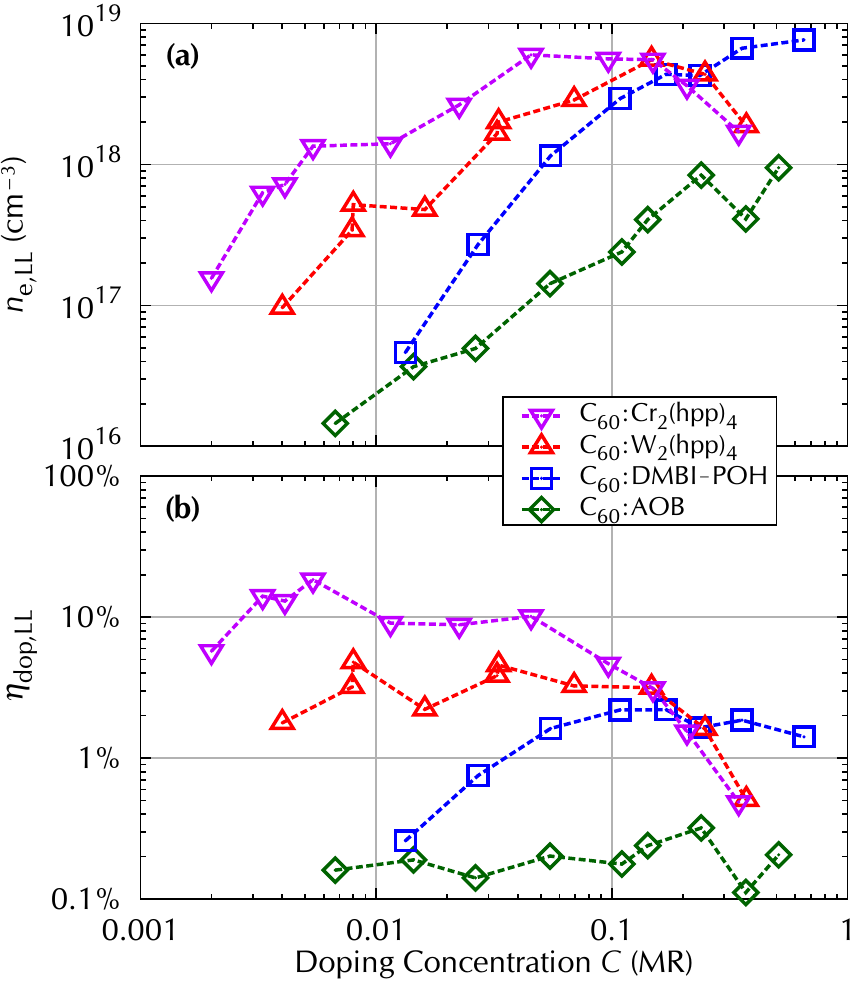}
{Lower limits of (a) \neLong \neLL and (b) doping efficiency \DopEffLL for $n$-doped \CS, calculated by assuming a constant mobility, set to the record value for intrinsic \CS of \mob[4.9]\cite{Itaka2006} and 
using the conductivity data from \figref{DataCond+See}\,(a), probed at \T[40].
}%

The derived \neLL values are depicted in \figref{calc-CCDLL+neLL-n}\,(a), calculated from the conductivity data shown in \figref{DataCond+See}\,(a). 
All material combinations show an increase of \neLL with \CLong. 
The highest \neLL close to $\SI{10^{19}}{\centi\meter^{-3}}$ are obtained for \CS highly doped by \dmbi, \CrPd or \WPd, whereas for \aob the largest value is one order of magnitude lower. These values have to be considered in relation to the total density of molecules of \mbox{$\nM{}_{\text{,C}_{60}}=\SI{1.36\cdot10^{21}}{\centi\meter^{-3}}$}. % calculated via \eqnref{nM-from-MatPara}.
For high concentrations of \CrPd and \WPd a saturation and decrease in \neLL is observed. 
It might possible that this is an artificial trend, produced by the assumption of a constant mobility. 
If the real values for \ne follow this trend, it is most probably originated in agglomeration and thus shielding of dopants. 

\X{In addition to} \neLL, the lower limit of the doping efficiency \DopEffLL for each sample is calculated using \eqnref{DopEffLL} and the results are presented in \figref{calc-CCDLL+neLL-n}\,(b). As mobility and doping efficiency are inversely proportional, the trends of the curves for \DopEffLL correspond to the trends of the lower limits of the mobilities \mobLL, presented in \figref{calc-MobLL-n}. \CS doped by \CrPd leads to a maximum value of \DopEffLL[18] at \C[0.005] and an almost constant \mbox{$\DopEffLL\approx10\,\%$} up to \C[0.045], followed by a decrease for the highest doping ratios used. Samples comprising \WPd yield efficiencies around \DopEffLL[3] and a drop at $\C>\mr{0.150}$.
\aob-doped samples have the lowest values of around \DopEffLL[0.2]. \dmbi samples start at a similar value, but rise 10-fold to a saturation around \DopEffLL[2] at high \CLongs. The gain in \DopEffLL for low \C of \dmbi samples is most probably correlated to an increasing real \DopEff in this range, as it is unlikely that only for one dopant the mobility of \CS is rising with the \CLong.

At low \CLongs, the real mobilities are expected to be least affected by the dopants and are consequently similar for all four material combinations.
Therefore, the calculated \DopEffLL at low \C can be directly compared and is expected to be correlated to the real doping efficiency \DopEff by a constant factor. This factor is given by the ratio of the used upper limit of the mobility \mobUL[4.9] and the real value of the bulk material in this sample geometry. Hence, at low doping concentration, the real doping efficiency \DopEff of \CrPd is approximately 3~times higher than for \WPd and around 15~times higher than for \aob and \dmbiPOH. Consequently, for low \C of these two dopants an upper limit $\DopEff{}_\text{UL} = 100\,\% \div 15 = 6.7\,\%$ is derived.

\subsection{Conclusions from Seebeck measurements}\label{sec:rechConclSeebeck}
\cBild
{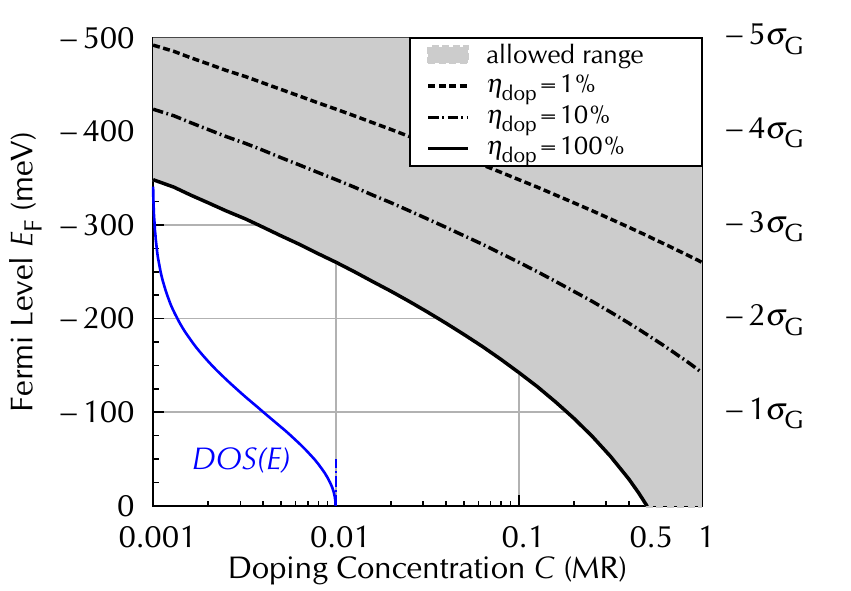}
{Calculated \EfLong position $\Ef(\C)$ with respect to the maximum of the Gaussian \dosLong for different doping efficiencies \DopEff. Derived using \eqnref{Ef-vonDopEff} and a Gaussian \dos with \gausswidth[100] and \T[40]. Only values in the gray area are physically allowed with $\DopEff\leq100\,\%$.
}
\begin{figure*}%
\centering
\includegraphics[width=\textwidth]{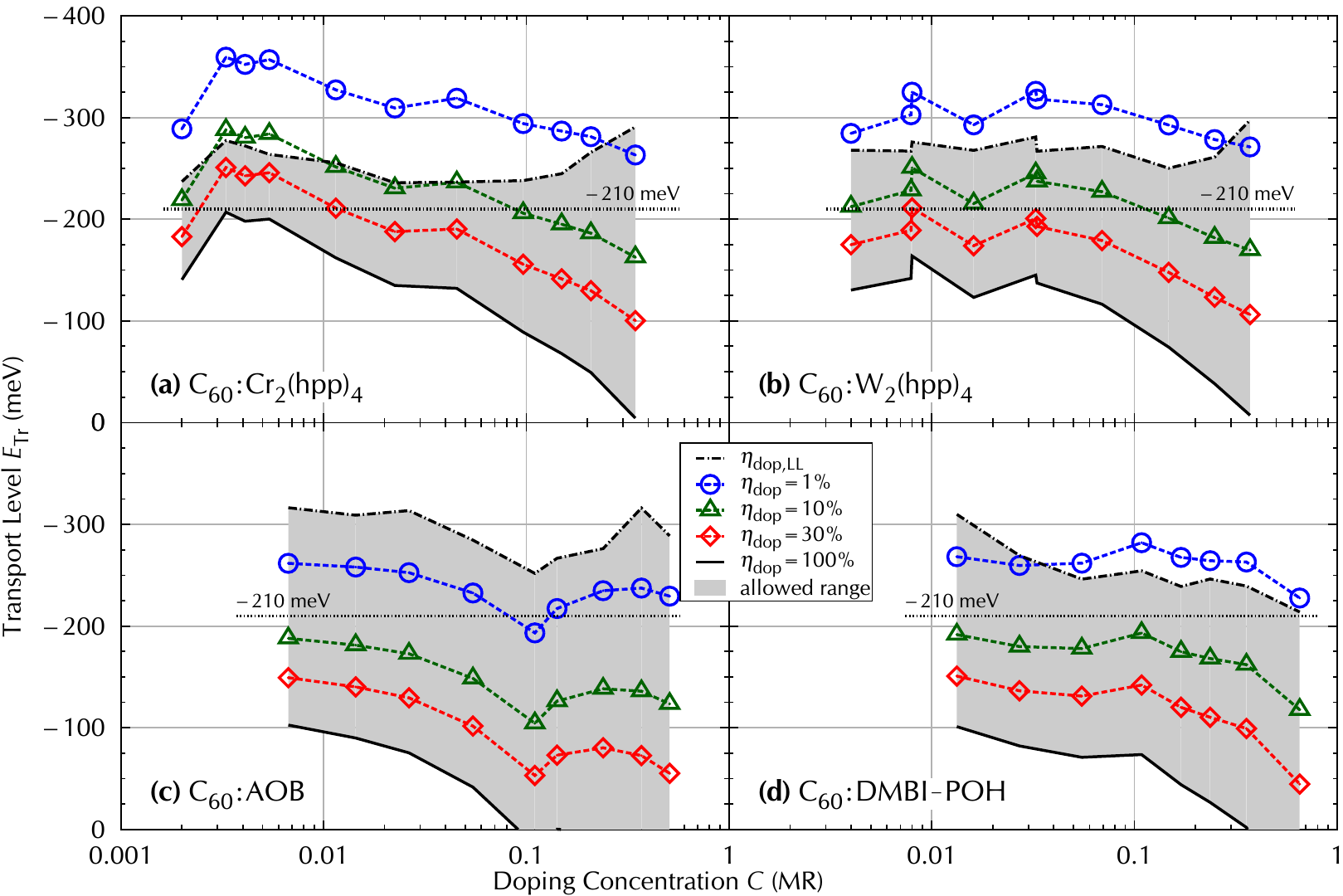}%
\caption{Calculated \EtLong position \Et with respect to the maximum of the Gaussian \dosLong for varying \CLong and \DopEffLong \DopEff. Obtained by subtracting measured \Es (\figref{DataCond+See}\,(b)) from calculated $\Ef(\C)$ (\figref{calc-Ef-von-DopEff}). Parameters used: \gausswidth[100] and \T[40]. The gray area corresponds to the physically allowed range between lower and upper limit of the doping efficiency. The dashed line at \Et[-210] indicates a value that is allowed for all samples.
}
\label{fig:calc-ETr-von-Es+DopEff}%
\end{figure*}
The \neLongL is furthermore given by the integral of the product of the \dosLong $\dos(E)$ and the \fFDLong $\fFD(E,\Ef)$ over all energies:
\begin{equation}
\ne = 
\int_{-\infty}^\infty \dos(E) \cdot \fFD(E,\Ef) ~ dE  \label{eq:CCD-basic-integral} 
\PUNKT
\end{equation} 
Thus, for a known $\dos(E)$ and given \DopEff, the position of the \EfLongL can be derived via comparing the \ne values with \eqnref{ne-from-DopEff-nD}.
Using a Gaussian density of states for modeling the energetic disorder of \CS and setting its maximum to the position at $E=0$, it follows:
\begin{align}
&\ne = \DopEff \cdot \nD \tag{\ref{eq:ne-from-DopEff-nD}}
\\
&=
\int_{-\infty}^\infty
\frac{\nH}{\sqrt{2 \pi} ~ \gausswidth} \exp{-\frac{E^2}{2\gausswidth^{2}}}
 \cdot
\frac{1}{1+\exp{\frac{E-\Ef}{\kT}}}
 ~ dE
\label{eq:Ef-vonDopEff} 
\PUNKT
\end{align}
Again, \nD and \nH are related to \C via equations~\eqref{C} and \eqref{nD}.
\gausswidth is the standard deviation of the distribution. In the following, a constant value of \gausswidth[100] for all \C is assumed, which is chosen to be somewhat higher than the reported \gausswidth[88]\cite{Fishchuk2010} for undoped \CS, to compensate the influence of doping that might broaden the \dos\cite{Pahner2013}.
The inversion of \eqnref{Ef-vonDopEff} to obtain \Ef as function of \C and \DopEff is performed numerically and the results are plotted in \figref{calc-Ef-von-DopEff}. 
The solid line represents the \Ef at \DopEff[100]. 
As the doping efficiency cannot exceed 100\,\%, only values above this line are physically allowed.
It can be seen that with increasing \C, \Ef reaches densely populated regions of the \dos, when assuming a constant \DopEff.

This approach is now combined with data from Seebeck studies, to calculate the position of the \EtLongL, with respect to the maximum of the \dos.
\Et is defined as the energy weighted by the differential conductivity $\c'(E)$\cite{Schmechel2003}
\begin{equation}
\Et
= \frac{1}{\c} \int_{-\infty} ^{+\infty} E ~ \c'(E) ~ dE
\PUNKT
\label{eq:DefEt}
\end{equation}
The measured Seebeck coefficient \S (at a certain \CLong) is directly proportional to the energetic difference between \EfLong and \EtLong\cite{Schmechel2003,Fritzsche1971}, denoted as \Es in the following.
\X{Subtracting this measured \Es from \Ef, the position of \mbox{$\Et=\Ef-\Es$} can be derived.}
\X{Again, \Ef is calculated as discussed above for a given \DopEff and varying \C. This calculation is performed for several different values of \DopEff, and the results are shown in \figref{calc-ETr-von-Es+DopEff}.}
As the doping efficiency \X{must be} be greater than the above derived lower limit \DopEffLL and cannot exceed \DopEff[100], only a certain region of \Et is consistent with all data, marked by the gray shaded areas in \figref{calc-ETr-von-Es+DopEff}. This physically possible region is for most samples between \Et[-300] and \meV{-100} with respect to the maximum of the Gaussian density of states. It is narrowest for \CrPd, due to the large value obtained for \DopEffLL.

\subsection{Assuming a constant transport level}\label{sec:rechConstEt}
\cBild
{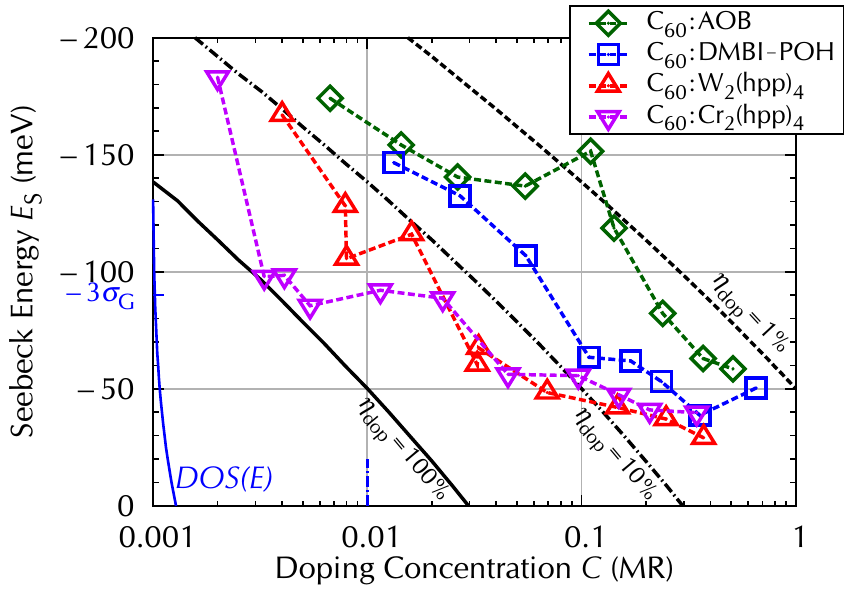}
{Measured \EsLong \Es (\T[40]) compared to calculated $\Es(\C)$ for constant \EtLong \Et[-210] at different doping efficiencies \DopEff. 
Calculations performed analogously to \figref{calc-Ef-von-DopEff} and subtracting \Et. Only values above the solid black line corresponding to \DopEff[100] are physically allowed. The \dos is sketched by the blue line using the same scale as in \figref{calc-Ef-von-DopEff}.
The \EfLong position \Ef is given by the sum of \Es and \Et.
}
\X{The narrow region of allowed \Et values suggests the assumption of a constant \EtLong for all samples and \CLongs as a further approximation. This allows for deriving values and general trends from the data.} A value of \Et[-210] is chosen, as this value is in the allowed regime for all samples, indicated by the dotted line in \figref{calc-ETr-von-Es+DopEff}.

Analogue to \figref{calc-Ef-von-DopEff}, trends for $\Es(\C,\DopEff)$ can be derived by subtracting this \Et[-210] from calculated $\Ef(\C)$, shown as lines in \figref{calc-Es-von-DopEff-210}.
These trends are now compared to the measured Seebeck data (symbols in \figref{calc-Es-von-DopEff-210}).
It can be seen that under the assumption of a fixed \Et, the Seebeck results of \CS doped by \Ckl{0.100} of \WPd or \dmbi follow the trend of a constant doping efficiency, whereas at larger \C the \Es tends towards lower \DopEff. The samples doped by \CrPd and \aob show deviations from the tendency of a constant \DopEff.

Using this fixed \Et[-210], the corresponding density of free electrons \ne is calculated for each sample by solving the integral \eqref{Ef-vonDopEff} of the product of \dos and Fermi distribution \fFD. Here, the measured \Es is used to calculate the \EfLong position $\Ef=\Et+\Es$. The results are shown in \figref{calc-constEtr-n-210}\,(a).

For all four material combinations, the calculated \ne increases with \C until at high \C a saturation is observed. Samples doped by \CrPd or \WPd saturate around \mbox{$\ne=\SI{10^{19}}{\centi\meter^{-3}}$} for \CLongs \mbox{$\C\geq\mr{0.040}$}. The same \ne is reached by \dmbi samples, but at higher \C.
\aob-doped samples saturate around lower \mbox{$\ne=\SI{5\cdot10^{18}}{\centi\meter^{-3}}$} for \Cgr{0.100}.
These values have again to be seen in relation to the density of molecules of \mbox{$\nM{}_{\text{,C}_{60}}=\SI{1.36\cdot10^{21}}{\centi\meter^{-3}}$}.
Overall, these trends seem to be more realistic than those derived under the assumption of constant mobility, compare \figref{calc-CCDLL+neLL-n}\,(a), where a decrease of the lower limit of the \neLong \neLL is found at high \CLongs of \CrPd and \WPd.

\cBild
{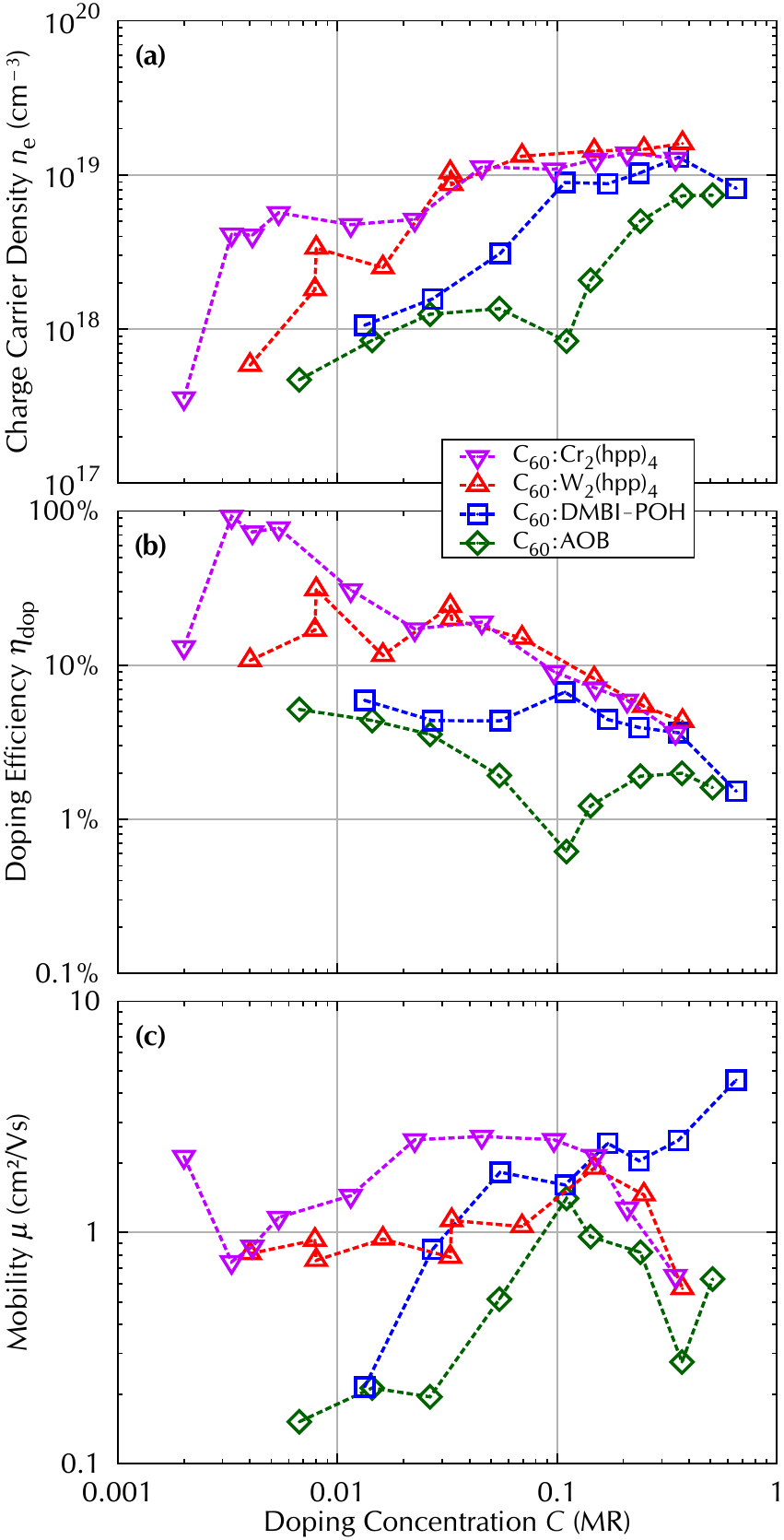}
{Calculated values of (a) charge carrier density \ne, (b) doping efficiency \DopEff and (c) mobility \mob for $n$-doped \CS. Based on the assumption of a constant \EtLong \Et[-210], using \gausswidth[100] and measured conductivity and Seebeck data shown in \figref{DataCond+See}.
}

The doping efficiency \DopEff can be derived 
from known \ne, as shown in \figref{calc-constEtr-n-210}\,(b). 
A maximum of \DopEff[92] is found for the sample of \C[0.0033] of \CrPd, showing that lower values than \Et[210] are not realistic, as these would result in an even larger value of \DopEff, which cannot exceed 100\,\%. Larger values of \Et on the other hand would lead to a violation of the lower limit of \DopEffLL, as derived in \secref{rechDopEffLL}. Hence, only values around \Et[-210] are compatible with all measurements. The doping efficiencies of \CrPd and \WPd samples decrease with \CLong and both series are in excellent agreement for $\C\geq\mr{0.040}$. \aob-doped samples show a similar trend but at lower values, whereas for \dmbi an almost \DopEff is observed.

The mobility \mob can be calculated from \eqnref{conductivity} by combining the derived values of the \neLong \ne and the measured conductivity data. The results are presented in \figref{calc-constEtr-n-210}\,(c). Rather high values are found, in agreement with the \mobLL, derived in \secref{rechMobLL}. Both, \CrPd and \WPd, yield an almost constant mobility at low and medium \C, followed by a decrease at high \C that might be attributed to changes in the morphology as discussed in ref.\cite{Menke2012}. Most of the mobilities derived for \WPd samples are lower than those for \CrPd.
This effect can be interpreted as doping by \WPd resulting into a reduction of the electron mobility, which might originate from its extremely small \IE $\text{IE}=\eV{2.68(13)}$ (compared to \eV{3.95(13)} for \CrPd)\cite{Menke2012} and thus strong tendency towards ionization. 
The samples doped by \aob or \dmbi show low mobilities at low \C and an increase in the medium doping regime. For \aob-doped samples, a decrease at high \C is observed, whereas for \dmbi the mobility rises further, up to a value of \mob[4.6], close to the expected limit of \mobUL[4.9].

Overall, the results derived on the basis of the assumption of a constant \EtLong \Et for all samples seem reasonable, as both, the values and the trends are in the expected range.
In general, it is expected that \Et, which is defined as the energy weighted by the differential conductivity $\c'(E)$ (compare \eqnref{DefEt}) shifts upon increasing \CLong towards the maximum of the Gaussian \dosLong (and hence to lower values), as the maximum of $\c'(E)$ is expected to shift in this direction. This would result in an upward shift of the trend of \ne and thus \DopEff with \C, whereas the mobility would be shifted downwards. Modeling this would require detailed knowledge on the energetic distribution of the mobility $\mob(E)$ contributing to $\c'(E)$\cite{Schmechel2003}.

\section{Conclusion}\label{sec:rechConclusion}
The simple models presented in the first part are powerful tools for deriving lower limits of the important parameters charge carrier mobility, \neLong and \DopEffLong from conductivity data of doped layers. These give an insight to the trends of the corresponding real values and allow to compare the relative values for different material combinations. The methods can easily be adopted for $p$-doped samples\cite{Menke2013,TorbenMenkeDiss} as well as for polymers.

Even without knowledge of the energetic dependency of the macroscopic mobility $\mob(E)$, it is shown that by combining Seebeck and conductivity studies, it is possible to narrow down the physically allowed regime for the \EtLong.

The assumption of a constant \EtLong position for all samples yields reasonable trends for \neLong, doping efficiency and mobility. %\todo{MKR: genauer?}
A more sophisticated model would require profound knowledge of the shape of the density of states and the energetic distribution of the mobility, as well as of the influence of doping on these, which are pathways for future studies.

\vspace*{0ex}
\section*{Acknowledgments} 
The authors thank Novaled GmbH, Germany for providing the dopants \CrPd and \WPd, 
and the Bao group at Stanford University, USA for providing \dmbi.

\end{document}